\documentclass[fleqn,10pt]{wlscirep}
\usepackage[utf8]{inputenc}
\usepackage[T1]{fontenc}
\usepackage{graphicx}
\usepackage{subcaption}
\usepackage[utf8]{inputenc}
\usepackage[export]{adjustbox}
\usepackage{wrapfig}
\usepackage{multicol}
\usepackage{blkarray}
\usepackage{amsmath}
\usepackage{ulem} 

\title{Adapting Modeling and Simulation Credibility Standards to Computational Systems Biology}

\author{Lillian T. Tatka${^1}$, Lucian P. Smith${^1}$, Joseph L. Hellerstein${^2}$ Herbert M. Sauro${^1}$}
\date{}
\begin{document}

\maketitle
\date{}

\begin{flushleft}
{\small 
${^1}$Department of Bioengineering, University of Washington, Seattle, WA, USA\\
${^2}$eScience Institute, University of Washington, Seattle, WA, USA}
\end{flushleft}

\flushbottom

\section*{Abstract}
Computational models are increasingly used in high-impact decision making in science, engineering, and medicine. The National Aeronautics and Space Administration (NASA) uses computational models to perform complex experiments that are otherwise prohibitively expensive or require a microgravity environment. Similarly, the Food and Drug Administration (FDA) and European Medicines Agency (EMA) have began accepting models and simulations as form of evidence for pharmaceutical and medical device approval. It is crucial that computational models meet a standard of credibility when using them in high-stakes decision making. For this reason, institutes including NASA, the FDA, and the EMA have developed standards to promote and assess the credibility of computational models and simulations. However, due to the breadth of models these institutes assess, these credibility standards are mostly qualitative and avoid making specific recommendations. On the other hand, modeling and simulation in systems biology is a narrow domain and several standards are already in place. As systems biology models increase in complexity and influence, the development of a credibility assessment system is crucial. Here we review existing standards in systems biology, credibility standards in other science, engineering, and medical fields, and propose the development of a credibility standard for systems biology models.

\begin{multicols}{2}

As computing power rapidly increases, computational models become more intricate and an increasingly important tool for scientific discovery. In systems biology, where the amount of available data has also expanded, computational modeling has become an important tool to study, explain, and predict behavior of biological systems. The scale of biological models ranges from subcellular components\cite{Wang2021} to entire ecosystems\cite{Hassell2021}. Modeling paradigms include mechanistic models, rule-based systems, Boolean networks, and agent-based models\cite{Bartocci2016}. This review will focus on mechanistic models of subcellular processes.

The Food and Drug Administration (FDA) defines model credibility as "the trust, established through the collection of evidence, in the predictive capability of a computational model for a context of use"\cite{FDAguidelines}. Model credibility is important in systems biology as models are used to guide experiments or to optimize patient treatment. This is particularly important given the increasing scale and intricacy of models. Reproducibility, the ability to recreate a model and data \textit{de novo} and obtain the same result\cite{shin_standards_nodate}, is directly connected to credibility, but even reproducibility remains a challenge. It was recently discovered that 49\% of published models undergoing the review and curation process for the BioModels~\cite{BioModels2020} database were not reproducible primarily due to missing materials necessary for simulation, the availability of the model and code in public databases, and lack of documentation \cite{tiwari_reproducibility_2021}. With some extra effort, an additional 12\% of the published models could be reproduced.  A model that cannot be reproduced is not credible. 

Due to the increasing importance of computational models in scientific discovery, the National Aeronautics and Space Administration (NASA), the Food and Drug Administration (FDA), and other regulatory bodies have developed standards to assess the credibility of models\cite{babula_nasa_2009,FDAguidelines,Shepard2015}. These standards are somewhat vague and generally qualitative to accommodate the broad scope of models in these fields. However, mechanistic models in systems biology are narrow in scope and are supported by a variety of standards for model encoding, annotation, simulation, and dissemination potentially enabling the development of a credibility standard for mechanistic systems biology models.

In this review, we discuss current systems biology modeling standards that could aid in the development of credibility standards, examine existing credibility standards in other scientific fields, and propose that current standards in systems biology and other fields could support the development of a credibility standard for mechanistic systems biology models.

\section{Current Standards in Systems Biology}

Klipp et al. describe standards as agreed-upon formats used to enhance information exchange and mutual understanding~\cite{Klipp2007StandardsIC}. In the field of systems biology, standards are a means to share information about experiments, models, data formats, nomenclature, and graphical representations of biochemical systems. Standardized means of information exchange improve model reuse, expandability, and integration as well as allowing communication between tools. In a survey of 125 systems biologists, most thought of standards as essential to their field, primarily for the purpose of reproducing and checking simulation results, both essential aspects of credibility~\cite{Klipp2007StandardsIC}.

A multitude of standards exist in systems biology for processes from annotation to dissemination. Although there is currently no widely used standard for model credibility, the development of this standard is likely to depend on existing systems biology standards, just as standards for model simulation are dependent on standards for model encoding. This section will summarize current standards relevant to model credibility including standards for ontology, encoding, simulating, and disseminating models.  Although standards also exist for graphical representation of systems biology models (SBGN)\cite{SBGN} and representation of simulation results (SBRML)\cite{SBRML}, these will not be discussed here as they are less relevant to the future implementation of model credibility standards. 

\subsection{Model Representation}
Having a commonly understood language for describing a model is essential in exchange, reproducibility, credibility. Without a common language to describe models, they cannot be simulated across different platforms or freely shared. For this reason, systems biology model representation has become standardized using XML-based languages SBML\cite{Klipp2007StandardsIC, machado_modeling_2011}, CellML\cite{CellML}, and BioPAX\cite{demir_biopax_2010}. NeuroML\cite{Gleeson2010}, similar to SBML and CellML, is used to represent neuronal models, but is beyond the scope of this review.  

\subsubsection{SBML}
The most widely used model format is SBML (Systems Biology Markup Language)\cite{SBML, Finney2003, Klipp2007StandardsIC, machado_modeling_2011}. SBML is a XML-based language for encoding mathematical models that reproduce biological processes, particularly biochemical reaction networks, gene regulation, metabolism, and signalling networks\cite{SBML, Kohl2011}. SBML encodes critical biological process data such as species, compartments, reactions, and other properties (such as concentrations, volumes, stoichiometry, and rate laws) in a standardized format. Annotations can also be stored in the SBML format. With its support by over 200 third party tools and its ability to easily convert to other model formats, SBML is the de facto language for systems biology models\cite{Kohl2011, machado_modeling_2011}.

SBML models are composed of entities, such as species, located in containers that can by acted upon by processes that create, destroy, or modify\cite{Keating2020}. Other elements allow for the definition of parameters, initial conditions, variables, and mathematical relationships. The SBML language is structured as a series of upwardly compatible levels, with higher levels incorporating more powerful features. Versions describe the refinement of levels. Most recently, SBML level 3 introduced modular architecture consisting of a set of fixed features, SBML level 3 core, and a scheme for adding packages that augment the core functionality. This allows for extensive customization of the language while enabling reuse of key features. Currently, eight packages are part of the SBML 3 standard. These packages extend the capability of SBML such as enabling descriptions of uncertainties in terms of distributions\cite{Smith2020}, allowing for the encoding, exchange, and annotation of constraint-based models\cite{Olivier2018},  rendering visual diagrams\cite{Gauges2015}, amongst many others\cite{Keating2020}.

\subsubsection{CellML}
Similar to SBML but broader in scope, CellML is also an XML-based language for reproducing mathematical models of any kind, including biochemical reaction networks\cite{Beard2009}. CellML models consist of several interconnected components\cite{Mesiti}. Each component must contain at least on variable and that variable must be associated with physical units. This enables CellML processors to automatically check equations for dimensional consistency. Mathematical equations describing how the component behaves within the model may also be included\cite{Wimalaratne2009}. For example, a component might represent a cell membrane and contain the variable for the potential difference across the cell membrane using voltage\cite{Mesiti}.

Both CellML and SBML use almost identical mathematical expressions in MathML, an international standard for encoding mathematical expression using XML\cite{Caprotti1999}. CellML explicitly encodes all mathematics, such as ODEs\cite{Smith2013}. It is more versatile than SBML, capable of describing any type of mathematical model. SBML defines reaction rates, which can be used to build rate rules and ODEs\cite{Hucka2015}. There is more third party support for SBML and it is a semantically richer language compared to CellML\cite{machado_modeling_2011, Smith2013}.

\subsubsection{BioPAX}
BioPAX (Biological Pathway Exchange) is an ontology, a formal systems of describing knowledge that structures biological pathway data making it more easily processed by computer software\cite{demir_biopax_2010}. It describes the biological semantics of metabolic, signalling, molecular, gene-regulatory, and genetic interaction networks \cite{demir_biopax_2010}. Whereas SBML and CellML focus on quantitative modelling and dynamic simulation, BioPAX concentrates primarily on quantitative processes and visualization\cite{demir_biopax_2010, buchel_qualitative_2012}.  

BioPAX contains one superclass, Entity. Within the Entity superclass, there are two main classes: PhysicalEntity and Interaction. PhysicalEntity describes molecules, including proteins, complexes, and DNA, while the Interaction class defines reactions and relationships between instances of the PhysicalEntity class. Interactions can be either Control or Conversion, both of which are divided into several more detailed subclasses\cite{demir_biopax_2010, buchel_qualitative_2012}. Like SBML, BioPAX is released level-wise with level 1 describing interactions, level 2 supporting signalling pathways and molecular interactions, and level 3 enabling the description of gene-regulatory networks and genetic interactions.

\subsection{Annotation}

As models grow more numerous and complex, there is an increasing need for a standardized encoding format to search, compare, and integrate them. While standards such as SBML and CellML provide information on the mathematical structure of a model, there is no information as to what variables and mathematical expressions represent. Simple textual descriptions of these representations are subject to errors and ambiguity and require text-mining for computational interpretation~\cite{Curtout2011}. Standardized metadata annotations capture the biological meaning of a model's components and describe its simulation, provenance, and layout information for visualization. Annotations improve model interoperability, reusability, comparability and comprehension~\cite{Gennari2021}. To avoid accounting for a variety of annotation formats and approaches, standard annotation protocols are necessary.
 
 Standardized protocols are enabled by systems biology specific ontologies~\cite{Wimalaratne2009}. Ontologies define a common vocabulary and set of rules to unambiguously represent information by identifying concepts, characteristics and specific instances of concepts, and describing ways in which these concepts and instances can relate~\cite{Noy2001OntologyD1}. In addition to annotation, ontologies can be useful to formally describe simulations, dynamics, and other aspects of systems biology models~\cite{Curtout2011}. Annotations improve model interoperability, reusability, comparability and comprehension~\cite{Gennari2021}. To avoid accounting for a variety of annotation formats and approaches, standard annotation protocols are necessary.
 
 Despite the numerous standards and tools, annotation remains a challenge. For example, the ChEBI database\cite{Hastings2015} has approximately 1,000 annotations for glucose. While more than one entry for each annotation can serve a purpose (some users may prefer to be more abstract in their annotations), this adds to the challenge of defining the purpose of a model and therefore its credibility.  Additionally, annotations can be obsolete, inappropriate or incorrect, or provide insufficient information. Evaluating the quality of annotations would be essential in any credibility assessment for systems biology models. Some tools already exist for this purpose, such as SBMate\cite{Shin2021}, a python package that automatically assesses coverage, consistency, and specificity of semantic annotations in systems biology models.

\subsubsection{MIRIAM}
 MIRIAM\cite{novere_minimum_2005} (Minimum Information Requested in the Annotation of Biochemical Models) was developed to encourage the standardized annotation of computational models by providing guidelines for annotation. The MIRIAM guidelines suggest that model metadata clearly references the relation documentation (e.g., journal article), that the documentation and encoded model have a high degree of correspondence, and that the model be encoded in a machine readable format (such as SBML or CellML). Annotations should also include the name of the model, the citation for its corresponding journal article, the contact information of the creators and date of creation, as well as a statement about the terms of distribution. Additionally, models should have accurate annotations that unambiguously links model components to corresponding structures in existing open access bioinformatics resources. The referenced information should be described using a triplet, {data collection, collection-specific identifier, optional qualifier} and expressed as a Uniform Resource Identifier (URI), a unique sequence of characters that identifies a resource used by web technologies\cite{berners2005uniform}. The optional qualifier field is used to describe relationships between the model constituents and the piece of knowledge with language such as "has a", "is a version of", "is homologous to", etc. 
 
\subsubsection{Systems Biology Ontology (SBO)}
SBO (Systems Biology Ontology) describes entities used in computational modeling\cite{SBO, Curtout2011}. It defines a set of interrelated concepts used to specify the types of components specified in a model and their relationships to one another. Annotation with SBO terms allows for unambiguous and explicit understanding of the meaning of model components and enables mapping between elements of different models encoded in different formats\cite{Curtout2011}. Both SBML and CellML support annotation with SBO terms.  SBML elements contain an optional sboTerm attribute\cite{Curtout2011, Hucka2015, Wimalaratne2009}.

\subsubsection{OMEX}
In order to harmonize the metadata annotations across models encoded in various formats, Gennari et al. with the consensus of the COMBINE (Computational Modeling in Biology Network) community developed a specification for encoding annotations in Open Modeling Exchange (OMEX)-formatted archives. The specification describes standards for model component annotations as well as for annotation at the model-level and archive-level \cite{Gennari2021}. The specification describes annotation best practices and addresses annotation issues such as composite annotations, annotating tabular data and physical units, as well as provides a list of ontologies relevant to systems biology. Implementation of these specifications is aided by LibOMexMeta, a software library supporting reading, writing, and editing of model annotations. It uses Resource Description Framework\cite{Decker2000} (RDF), an XML-based standard format for data exchange on the web, for representing annotations. It also makes use of several standard knowledge resources about biology and biological processes such as ChEBI\cite{Hastings2015}, a dictionary of small chemical compounds, and UniProt\cite{Apweiler2004}, a database of protein sequence and functional information.

\subsubsection{Annotation in CellML and SBML}
 Both CellML and SBML have their own annotation protocols based on RDF\cite{Wimalaratne2009}. The CellML language uses its own ontology for model annotation, a necessity due to the flexibility of the language\cite{Beard2009}. The CellML Metadata specification was developed parallel to the CellML language \cite{Wimalaratne2009}. CellMLBiophysical/OWL ontology is composed of two categories: physical and biological\cite{Wimalaratne2009}. The physical ontology describes physical quantitative information and concepts captured in the model's mathematical expressions. It is subdivided into processes, such as enzyme kinetics, ionic current, and rate constants, and physical entities, such as area, concentration, volume, and stoichiometry. The biological ontology provides description for processes, entities, the role of an entity in relation to a process, and the specific location of the entity in a biological system. Bioprocesses are divided into three subclasses: biochemical reactions, transport, and complex assembly. Biological entities include protein, small molecule, and complex. The biological roles subclass is composed of modifier, reactant, and product. 

SBML also facilitates MIRIAM compliant annotation using the resource description framework (RDF)\cite{Swainston2009,Decker2000}. Annotations use BioModels.net\cite{LeNovere2006} qualifier elements embedded in XML form of RDF\cite{Hucka2019}. Each annotation is a single RDF triple consisting of the model component to annotate (subject), the relationship between the model component and the annotation term (predicate), and a term which describes the meaning of the component (object). These terms come from defined ontologies, such as SBO\cite{SBO}. RDF annotation is supported by the software libraries libSBML\cite{libSBML} and JSBML\cite{Rodriguez2015}.

\subsection{Simulation and Parameter Estimation}
Information about a model alone is insufficient to enable efficient reuse. A variety of advanced numerical algorithms and complex modeling workflows make the reproduction of simulations challenging. Many modelers reproduce simulations by reading the simulation description in the corresponding publication\cite{waltemath_minimum_2011}. This is time consuming and error prone and often, the published description of a simulation is incomplete or incorrect. For these reasons, it is essential to define and include information necessary to perform all simulations.

\subsubsection{MIASE}
Guidelines for the Minimum Information About a Simulation Experiment (MIASE) were introduced to specify what information should be provided in order to correctly reproduce and interpret a simulation\cite{waltemath_minimum_2011}. MIASE is a set of rules that fall into three categories: information about the model used in the simulation experiment must be listed in a way that enables reproduction of the experiment; all information necessary to run any step of the experiment must be provided; all information needed to post-process data and compare results must be included.  Along with MIRIAM\cite{novere_minimum_2005} guidelines, MIASE compliance guarantees that the simulation experiment is true to the intention of the original authors and is reproducible. 

\subsubsection{KiSAO}
KiSAO (Kinetic Simulation Algorithm Ontology) is an ontology used to describe and structure existing simulation algorithms\cite{Zhukova2011, Curtout2011}. It consists of three main branches, each with several subbranches. The first branch is Kinetic simulation algorithm characteristics such as the type of system behavior or type of solution. The second in the kinetic simulation algorithm such as Gillespie or accelerated stochastic simulation. The third branch is kinetic simulation algorithm parameters which describes error and granularity, amongst other characteristics. 

\subsubsection{SED-ML}
Simulation Experiment Description Markup Language (SED-ML) is software independent, XML-based format for encoding descriptions of simulation experiments and results\cite{bergmann_simulation_2018, Smith2021simulation}. To help modelers comply with MIASE rules, SED-ML describes the details of simulation procedures including what datasets and models to use, which modifications to apply to models, which simulations to run on each model, how to post-process data, report, and present results can all be encoded\cite{waltemath_minimum_2011}. Each algorithm mentioned in a SED-ML file must be identified by a KiSAO term\cite{Curtout2011}. PhraSED-ML was developed to enable modelers to encode human readable SED-ML elements without the use of specialized software\cite{choi_phrased-ml_2016}.

\subsubsection{PEtab}
Parameter estimation is common in modeling and simulation, which often requires running several simulations to scan the suitability of several parameter sets.  Although several parameter estimation toolboxes exist, they each use their own input formats. The lack of a standardized format makes it difficult to switch between tools, hindering reproducibility\cite{Schmiester2021}.  PEtab is a parameter estimation problem definition format consisting of several files containing information necessary for parameter estimation including the model (in SBML format), experimental conditions, observables, measurements, parameters, and optional visualization files\cite{Schmiester2021}. A final PEtab problem file links all other files to form a single, reusable, parameter estimation problem. Following the success of PEtab, parameter estimation functionality was added to SED-ML\cite{Smith2021simulation}.

\subsection{Dissemination}
Model reproducibility best practices describe dissemination as an essential part of reproducibility\cite{porubsky_best_2020}. Sharing all model artifacts and documentation on an open-source repositories allows independent researchers to reproduce, reuse, and understand the model. Several guidelines and archive formats have been developed to ensure that all relevant information necessary to reproduce a modeling result is easily accessible to the public.

\subsubsection{MIRIAM Curation Guidelines}
In addition to annotation guidelines, MIRIAM also provides guidelines for model curation, the process of collecting and verifying models. The aim of MIRIAM guidelines is the ensure that model is properly associated with a reference description (e.g. a journal article) and that it is consistent with that reference description. The model must be encoded in a public, machine-readable format such as SBML or CellML and comply with the standard in which it is encoded. The model must be related to a single reference description and reflect the biological process listed in the reference description. The encoded model must be simulatable, including quantitative values for initial conditions, parameters, and kinetic expressions and must reproduce relevant results when simulated\cite{novere_minimum_2005}.

\subsubsection{FAIR}
More recently, the FAIR guidelines were published to improve the ability of computers to \textbf{F}ind, \textbf{A}access \textbf{I}nteroperate, and \textbf{R}euse models\cite{Wilkinson2016} with minimal human interaction. FAIR differs from most data management and archival guidelines in that it is a set of high-level, domain independent guidelines that can be applied to a variety of digital assets. Each element of the FAIR principles is independent. They define characteristics that data resources should possess to assist with discover and reuse by third-parties. 

For a model to be "findable," it should be easy to find for both humans and computers. This requires describing and annotating data and meta data with unique identifiers that are registered or index in a searchable resource. Once the user finds the relevant model, it should be accessible: data and metadata should be retrievable by their identifiers using standard communications protocol. Metadata should remain accessible even when data are no longer available. Interoperability refers to the integration with other data and the ability to operate with various applications and workflows. This is enabled by the use of broadly applicable languages for model representation and annotation. The ultimate goal of FAIR is to enable the reuse of data. Data and metadata should be associated with detailed provenance, meet domain-specific community standards (such as COMBINE archive format described below), and released with clear and accessible data usage license.

\subsubsection{COMBINE Archives}
COMBINE (COmputational Modelling in BIology NEtwork) is a formal entity that coordinates standards in systems biology.  To assist in this coordination, a MIRIAM compliant system for sharing groups of documents regarding a model was developed called the COMBINE Archive\cite{schreiber_specifications_2020}. The archive is encoded in OMEX (Open Modeling Exchange format) and the archive itself is a "ZIP" file. A COMBINE archive could contain files in several different standard formats including SBML, SBOL, and SED-ML amongst others. Additionally, every COMBINE Archive contains at least one file called \textit{manifest.xml} that contains a list of all the files comprising the archive and describing their locations. An archive also may contain a metadata file, ideally conforming to MIRIAM and MIASE guidelines. The inclusion of all necessary protocols and data needed to implement a model enables distribution of models via a single file encouraging reuse and improving reproducibility\cite{bergmann_combine_2014}.

\section{Credibility Guidelines in Systems Biology}

Although no standard for model credibility in systems biology exists, there are general guidelines aimed at improving the trustworthiness of models, developed by the Committee on Credible Practice of Modeling and Simulation in Healthcare, formed by the U.S. National Institutes of Health\cite{erdemir_credible_2020} to enable the credible use of modeling and simulation in healthcare and translational research. These guidelines are qualitative and share many components with best practices for reproducibility. The term "credible" was defined as  "dependable, with a desired certainty level to guide research or support decision-making within a prescribed application domain and intended use; establishing reproducibility and accountability." These guidelines are qualitative and intended to cover a variety of modeling approaches and applications within the biomedical context. 

The credibility of a model should be evaluated within the model's context of use\cite{erdemir_credible_2020}.
 To this end, the guidelines recommend using contextually appropriate data and evaluating the model (performing verification, validation, uncertainty quantification and sensitivity analysis) with respect to the context in which the model will be used. Any limitations should be listed explicitly. 

Borrowing from software engineering best practices, the guidelines also recommend the use of version control to track model and simulation development as well as extensive documentation of simulation code, model mark-up, scope and intended use. Models should also include guides for developers and users and conform to domain-specific standards \cite{erdemir_credible_2020}.

Different simulation strategies should tested to ensure that the results and conclusions are similar across various tools and methods. All modeling components such as software, models, and results should be reviewed by third party users and developers and disseminated widely.

\begin{center}
    \captionsetup{type=figure}
    \includegraphics[width=6cm]{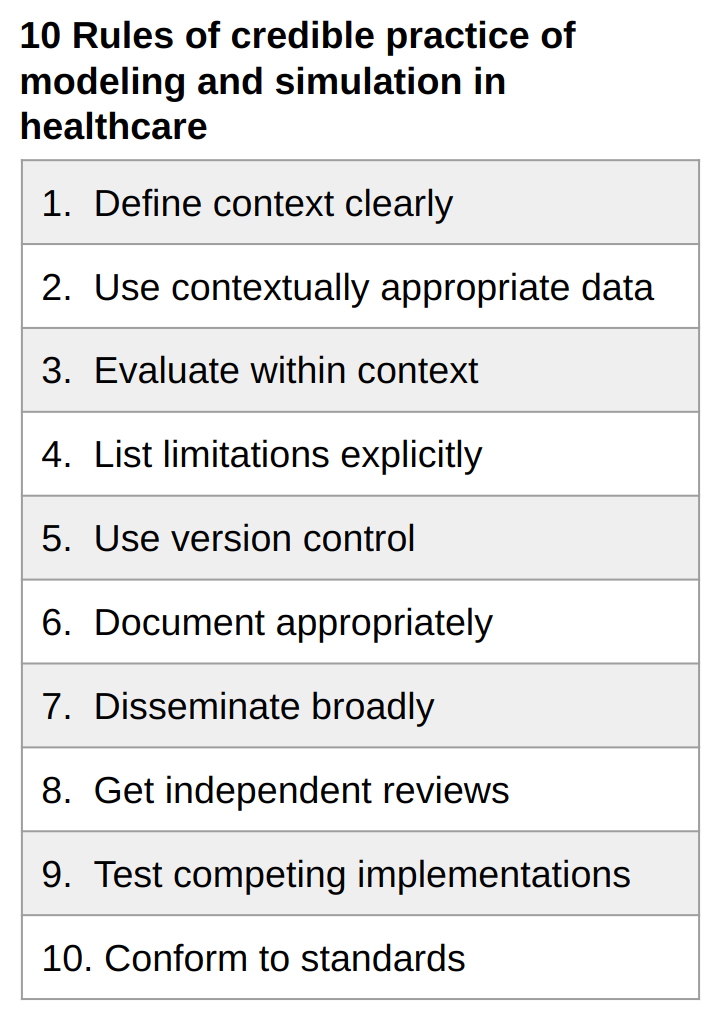}
    \captionof{figure}{The Committee on Credible Practice of Modeling and Simulation in Healthcare 10 rules of model credibility.}
    \label{fig:10rules}
\end{center}

\section{Qualitative Credibility Assessment in Other Modeling Fields}
National Aeronautics and Space Administration (NASA) and the Food and Drug Administration (FDA) also have a keen interest in producing well-documented and credible models for the purpose of making critical decisions. However, modeling and simulation tasks in these institutions are far broader compared to systems biology. NASA models range from the analysis of individual parts to orbits and spacecraft while models submitted to the FDA include medical devices and pharmacokinetics. Due to the wide variety of modeling tasks in NASA and the FDA, credibility guidelines in these institutions are general, largely qualitative, and do not prescribe specific tests.

\subsection{NASA Standard for Models and Simulations: Credibility Assessment Scale}

After the loss of the Columbia Space Shuttle and its seven crew members in 2003, the U.S. National Aeronautics and Space Agency (NASA) significantly increased its focus on quantitative and credible models. The misuse of an existing model and the reliance on engineer's judgement led to the false conclusion that shuttle reentry would not be affected by a small hole in the heat caused by a debris strike during takeoff\cite{Blattnig2013-qx, Howell2021}. The lack of quantifiable uncertainty and risk analysis in the report to management ultimately led to the shuttle's disintegration\cite{Niewoehner2008}. Since then, NASA has developed extensive modeling and simulation standards including the Credibility Assessment Scale\cite{babula_nasa_2009} (CAS) (\ref{fig:NASA}.  

Each model credibility standard described here emphasizes assessments be made within a specific context of use (COU), the specific role and scope of the model and the specific question of interest that the model is intended to help answer\cite{FDAguidelines}. The judgement error that ultimately led to the Columbia Space Shuttle disaster was partially due to the use of a modeling software far outside the intended context of use, leading to incorrect predictions and over-reliance on engineer's judgement\cite{Blattnig2013-qx}. In addition to specifying the scope and question of interest, the context of use should also describe how model outputs will be used to answer the question of interest and whether other information, such as bench-testing, will be used in conjunction with the model to answer the question of interest\cite{FDAguidelines}. The standards described here, from various institutions such as NASA and the FDA, all specify that credibility is to be evaluated within a specific context of use.

\begin{center}
    \captionsetup{type=figure}
    \includegraphics[width=8.5cm]{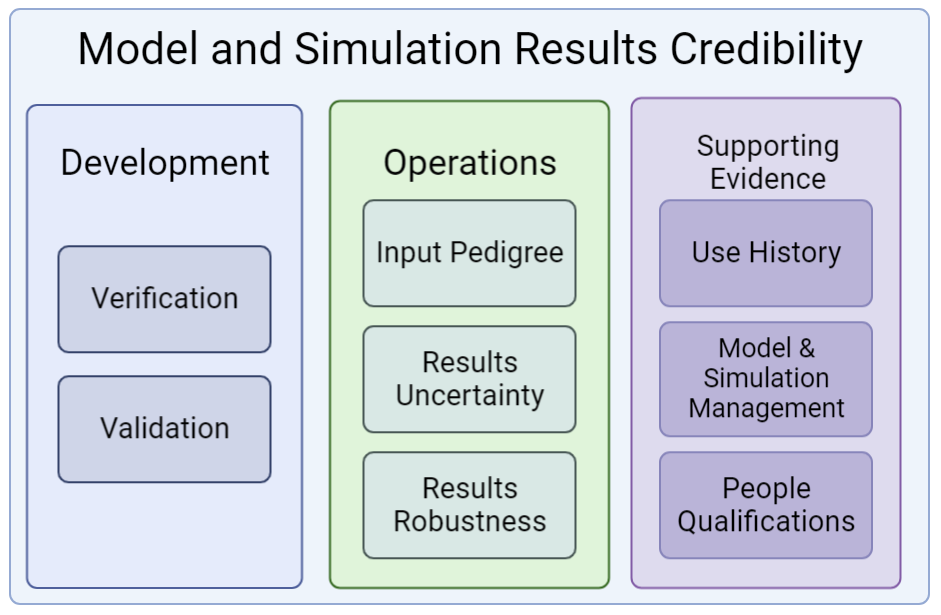}
    \captionof{figure}{Categories of the NASA Credibility Assessment Scale (CAS)}
    \label{fig:NASA}
\end{center}

NASA's Credibility Assessment Scale (CAS) is intended to help a decision-maker evaluate the credibility of specific modeling and simulation results and to identify aspects of the results that most influence credibility\cite{blattnig_towards_2008, Blattnig2013-qx}. The credibility assessment process can be viewed as a two part process: first the modeler conveys an assessment of the results, then a decision maker infers the credibility of these results. The CAS standard consists of eight factors grouped into three categories\cite{babula_nasa_2009}: development, operations, and management. Each of the eight factors is scored on a scale of 0-4 with guidelines for each numeric score. These factors were selected as they were considered to be essential factors that were sufficiently independent of one another and could be assessed objectively. While the primary concern is the score for each individual factor, the secondary concern is the score of the overall model, which is the minimum score of the eight subfactors. 

The model and simulation (M\&S) development category consists of subsections verification and validation\cite{babula_nasa_2009}. Scoring in these subcategories assess the correctness of the model implementation, the numerical error and uncertainty, and the extent to which the M\$S result matches reference data. If numerical errors for important features are "small" and if results agree with real-world data, the highest score of 4 is awarded for these factors.

The second category, M\&S operations, consists of three factors: input pedigree, results uncertainty, and result robustness\cite{babula_nasa_2009}. Input pedigree describes the level of trust in the input data, where input data that accurately reflects real-world data receiving the highest score. The results uncertainty category earns the highest score if non-deterministic numerical analysis (or is it non-deterministic and numerical analyses are) is performed. Result robustness high scores are achieved by including sensitivity analysis for most parameters and key sensitivities are identified.

Model and simulation management, the third category, is less technical, containing the factors use history, M\&S management, and people qualification\cite{babula_nasa_2009}. Use history scores the highest score if the model has previously been used successfully and meets "de facto" standards. For example, a model used for finite element analysis (FEA) would be required to meet FEA standards and codes for the type of object being modeled. M\&S management refers to the maintenance and improvement of the model with continual process improvement receiving the highest score of 4.  The people qualification category assesses the experience and qualifications of those constructing, maintaining, and using the model where personnel with extensive experience with the model and best practices scoring the highest. 

Although these categories were chosen in part by their ability to be objectively assessed, there is still a significant subjective component of the scoring process. It is acknowledged that different decision makers may assign different degrees of credibility to the same model and different decisions may require different levels of credibility. The CAS serves as a template to assess and clearly communicate risks to decision-makers. Additionally, it can be useful in measuring model development progress or in identifying areas where improvement is most needed\cite{Blattnig2013-qx}.

\subsection{Credibility Standards for Medical Models}
In addition to systems biology, computational models are also becoming essential tools in biomedical applications such as drug discovery\cite{Chen2015}, pharmacokinetics\cite{Kuh2000}, and medical devices\cite{Kung2019}. Credibility is essential in biomedical modeling, particularly in cases where models influence patient treatment or regulatory approval of a device or drug. Both the FDA and the European Medicines Agency have developed standards guiding model credibility for the purposes of regulatory approval. As with NASA, these guidelines are broad and qualitative due to the broad scope of biomedical modeling. Before the FDA began formalizing guidelines for model credibility the American Society of Mechanical Engineers (ASME) issued Verification and Validation (V\&V) 40 for assessing credibility of computational modeling in medical device applications\cite{viceconti_credibility_2020}. This standard assumes the ability to perform traditional validation activities such as comparing model predictions to well-controlled validation experiments\cite{FDAguidelines}. The FDA recognized that models used in regulatory submissions are often supported by many sources of evidence such as clinical trials or population-level validation. For this reason, FDA modeling credibility guidelines rely on ASME V\&V 40 concepts, but provide a more general framework for assessing a wider variety of models.

\subsubsection{ASME V\&V 40}

The American Society of Mechanical Engineers (ASME) developed the verification and validation standard 40 (V\&V 40) in 2012 as a standard describing verification and validation activities in the modeling and simulation of medical devices\cite{viceconti_credibility_2020}. Like the NASA CAS, V\&V focuses on context of use, model risk, and the establishment of credibility goals prior to any credibility assessment. The context of use addresses the specific role of the model in addressing the question of interest. Model risk is then assessed based on the possibility that the model may lead to incorrect conclusion resulting in adverse outcomes. After the establishment of credibility goals, verification and validation take place. The parts of this process especially relevant to modeling in systems biology are code and calculation verification.

Verification seeks to determine if the systems is being built correctly. Code verification aims to identify any errors in the source code and numerical algorithms. This can be done by comparing output from model to benchmark problems with known solutions\cite{viceconti_credibility_2020}. Calculation verification estimates the error in the output of a model due to numerical methods. Output errors can include discretization errors, rounding errors, numerical solver errors, or user errors. Calculation verification is complete when it is demonstrated that errors in the numerical solution are minimized to the point that they are not corrupting the numerical results\cite{viceconti_credibility_2020}.

Validation assesses how well the computational model represents reality. This might include comparing the model's behavior to the biological features of the real phenomenon by comparing results to \textit{in vitro/in vivo} benchmark experiments. Validation also includes uncertainty quantification and sensitivity analysis. Uncertainty quantification refers to the estimation of how stochastic error in the input propagates into the model output and sensitivity analysis is a post-hoc examination of the results of the uncertainty quantification to evaluate which elements most influence output variability\cite{viceconti_credibility_2020}.

Unlike the NASA CAS, V\&V 40 does not describe the quality of evidence needed to prove a model credible and lacks an objective scoring system necessary for implementing "cut-offs" of credible versus non-credible models, or for comparing the credibility of multiple models.

\subsubsection{FDA Guidance on Computational Model Credibility in Medical Devices}

Based on the V\&V 40 standard, the FDA recently released guidance on assessing credibility for models of medical devices\cite{FDAguidelines}. This guidance extends V\&V 40 by accepting other forms of evidence to demonstrate credibility in addition to traditional verification and validation exercises. Applicable to physics-based, mechanistic, or other first-principles-based models of medical devices, these guidelines consist of ten categories broadly divided into code verification, calculation verification, and validation. The code verification category is taken directly from V\&V 40.

The calculation verification guideline extends the V\&V 40 by detailing several methods to verify that the model is producing the intended output.  The model results can be compared with the same data used to calibrate the model parameters. Broader evidence in support of the model, but perhaps without a specific context of use are also acceptable. A model can also be verified using \textit{in vitro} or \textit{in vivo} experiments either within the context of use, or within conditions supporting a different context of use. These techniques can also be used for validation evidence. 

Validation assesses the model's ability to reproduce real-world behavior. In addition to the methods described for calculation verification, validation can also include population-based evidence, statistical comparisons of model predictions to population-level data such as the results of a clinical trial. Credibility is also supported by emergent model behavior, the ability of a model to reproduce real-world phenomena that were not pre-specified or explicitly modeled, as well as general model plausibility, that model assumptions, input parameters, and other characteristics are deemed reasonable based on scientific knowledge of the system modeled. 

Unlike the NASA Credibility Assessment Scale, these FDA guidelines are sets of nonbinding recommendations. Additionally, no scoring or suggested quality measures of FDA credibility factors are included making quantitative analysis of credibility impossible.

\subsubsection{EMA Guidelines for PBPK Models}

Of particular relevance for the field of systems biology is The European Medicines Agency's (EMA) Guideline on the Reporting of Physiologically Based Pharmacokinetic (PBPK) Modelling and Simulation issued in 2018\cite{viceconti2021, Shepard2015} . PBPK models are mathematical models that simulate the concentration of a drug over time in tissues and blood. With the rise in regulatory submissions that include PBPK models that rely on specialized software programs, this guidelines provides detailed advice on what to include in a PBPK modeling report. 

The standard describes information needed to describe and justify model parameters. Like the FDA standard, modelers are required to submit any assumptions made when assigning parameters and to document the sources of any literature-based parameters. Additionally, modelers must perform a sensitivity analysis for parameters that are key to the model (those that significantly influence the outcome) and list any parameters that are uncertain. 

The submission must include the simulation results as well as the relevant files used to generate the final simulations in both tabular and executable format. This requirement is shared with reproducibility standards already in place for systems biology in the COMBINE Archive standard as well as described in systems biology modeling reproducibility best practices\cite{porubsky_best_2020}.

The predictive performance of the model must also be evaluated. That is, its ability to recapitulate observed pharmacokinetics. This requirement is also mentioned in the FDA guidelines. 

Lastly, a discussion of confidence in the model and how potential uncertainty may influence it is required. Although this requirement is described more qualitatively in the EMA standard, the requirement that quantitative uncertainty be included can also be found in NASA's CAS, V\&V 40, FDA guidelines, as well as in reproducibility best practices for systems biology models\cite{porubsky_best_2020}.

\subsection{Current Tools for Systems Biology Model Testing}
Although there is no credibility standard in systems biology modeling, some tools provide automated model testing. Although these tools were not developed explicitly to assess credibility, many of the factors they test for could be considered aspects of credibility. Future model credibility assessments could aspire to the quantitative nature and automation these tools offer. 
 
\subsubsection{MEMOTE}
MEMOTE (MEtabolic MOdel TEsts) is an open-source Python software that automatically tests and scores genome-scale metabolic models (GEMs)\cite{Lieven2020-kj}. MEMOTE offers a web interface where SBML files can be uploaded and analyzed and ultimately scored. The tests check that a model is annotated according to the MIRIAM standard, that components are described using SBO terms, an that the model is properly constructed using the relevant SBML package, SBML-FBC\cite{Lieven2020-kj, Olivier2018-tx} . Basic tests check for the presence of relevant components, charge information, and metabolite formulas. Biomass tests check that biomass precursors are produced and that growth rate is non-zero. Stoichiometry tests test for inconsistency, erroneously produced energy metabolites, and reactions that are permanently blocked. A numeric score is output after testing indicating the extent to which a model conforms to these standards.

MEMOTE is designed to assess genome-scale metabolic models and largely includes tests that are specific to this model subset. Although a high MEMOTE score is likely to be indicative of model quality and reproducibility, it is not an assessment of credibility. A credible model will likely have a good MEMOTE score, but a good MEMOTE score does not necessarily indicate a credible model. However, the quantitative and automated nature of MEMOTE allows for quickly gauging model quality, comparing models, and the iterative improvement of metabolic models.

\subsubsection{FROG Analysis}
Similar to MEMOTE, the COMBINE community has recently developed FROG analysis, an ensemble of analyses for constraint-based models to generate standardized numerically reproducible reference datasets\cite{sbmlsim21}. Results from constraint-based models are often communicated as flux values and there are often multiple solutions for a single model. As such, results cannot be used to gauge reproducibility. The COMBINE community outlined a list of outputs and results of flux balance analysis (FBA) that are numerically reproducible and can be used for curation, known as FROG reports. FROG reports can be used in the BioModels\cite{BioModels2018a, BioModels2020} curation process to assess reproducibility.

FROG analysis consists of  \textbf{F}lux variability analysis, \textbf{R}eaction deletion, \textbf{O}bjective function values, and \textbf{G}ene deletion fluxes. Flux variability analysis (FVA) test that the maximum and minimum fluxes are reproducible using different software tools. The objective function value for a defined set of bounds should be reproducible. The systematic deletion of all reactions or all genes, one at a time, should provide comparable reference results. Currently four tools support the generation of FROG reports. Web-based tools include fbc\_curation\cite{sbmlsim21}, CBMPy MOdel Curator\cite{SBMpy} (both of which are also available as command line tools), and FLUXER\cite{hari2020}. fbc\_curation\_matlab is a command-line tool and exports results in COMBINE archive format\cite{fbc_curation}.

Unlike MEMOTE, FROG analysis produces a report in lieu of a single numerical score. 

\section{Discussion}
When standards are established, tests can be established to assess the extent to which a model conforms to that standard. With the development of standardized quantitative metrics (as opposed to qualitative guidelines such as those discussed previously), models can be constructed to meet minimum quality requirements lending credibility to those models and allowing for easy comparison across models\cite{Kaddi2007-js}. 

The difficulty in developing these quantitative metrics is that the characteristics of an ideal bio-model must be known and expressed concisely. Existing standards in systems biology seek to address the first point by outlining what information is necessary to completely define and reproduce a model as well as the format in which that information is to be presented. However, a model could meet all existing standards and not be credible. For example, a model could be fully defined in SBML with extensive annotations, be reproducible, properly formatted for dissemination with SED-ML files describing all simulations. Despite meeting these standards, this hypothetical model could produce negative concentrations when simulating, clearly indicating that the model is not credible.  Additional metrics  and standards must be established to adequately assess credibility. These metrics might include the relative concentration of floating species or the shape of response curves.

Hellerstein et al. note that several issues in biomedical modeling are analogous to problems faced in software development and propose that software development best practices might be translated to improve modeling in systems biology\cite{Hellerstein2019}. Of particular interest is software testing, which can be considered a form of credibility assessment. These tests aim to ensure the correctness, reliability, and availability of software, all characteristics that are also essential in systems biology model credibility.

Software tests can be divided into two categories, which may also be applicable in systems biology modeling: (i) black-box testing and (ii) white-box testing. Black-box software testing assesses the behavior of the code and does not deal with implementation. For systems biology model credibility assessment, black-box credibility indicators might be that the model accurately predicts observed data. White-box testing evaluates the internal workings of a software project or model. The absence of errors, such as undefined parameters, typos, or un-used species, might serve as white-box credibility indicators.

\section{Conclusion}

Although many reproducibility standards are in use to simplify assessing reproducibility, there are no standards and scoring systems for model credibility in systems biology. Unlike institutions such as NASA and the FDA, which deal with models spanning a broad scope of applications and scales, systems biology is focused on the modeling of cellular processes. This narrow scope, combined with the variety of standards already in use, makes systems biology models well-suited for a credibility standard.

A quantitative credibility scoring system would be particularly useful and enable comparing credibility of different models and guide the development of more credible models. Credibility metrics could be published along side models to indicate the trustworthiness of results and allow users to make informed decisions about reusing models.

Systems such as MEMOTE demonstrate that model standards and model quality indicators can be automatically quantitatively scored enabling iterative improvement during the development phase. More challenging is further developing standards to express characteristics, both quantitative and qualitative, that make a model credible. Current modeling standards in other scientific fields emphasize assessing credibility in the model's context of use. This poses a challenge for automating credibility assessment in systems biology modeling as more or less rigor may be required to achieve a sufficiently credible model depending on the intended use of the model. It may prove useful to develop a manual, semi-quantitative scoring systems, such as NASA's Credibility Assessment Scale (CAS) prior to attempting to implement a fully quantitative and perhaps automated credibility scoring system for systems biology models.

\end{multicols}

\section*{Funding}
This work was supported by NIH Imaging and Bioengineering (NIBIB) award P41GM109824, and the National Science Foundation award 1933453. The content expressed here is solely the responsibility of the authors and does not necessarily represent the official views of the National Institutes of Health, the National Science Foundation, or the University of Washington.

\bibliography{researchbib}
\end{document}